\documentclass[preprint,showpacs,preprintnumbers,showkeys]{revtex4}
\usepackage{amsfonts}
\usepackage{amsmath}
\usepackage{amssymb}

\begin{document}

\title{Cosmological Equations and Thermodynamics on Apparent Horizon in
Thick Braneworld}
\author{Shao-Feng Wu$^{1,2}$\footnote{%
Corresponding author. Email: sfwu@shu.edu.cn; Phone: +86-021-66136202.},
Guo-Hong Yang$^{1,2}$\footnote{%
Email: ghyang@mail.shu.edu.cn}, and Peng-Ming Zhang$^{3,4}$\footnote{%
Email: zhpm@impcas.ac.cn}} \affiliation{$^{1}$Department of physics,
Shanghai University, Shanghai, 200436, P. R. China}
\affiliation{$^{2}$The Shanghai Key Lab of Astrophysics, Shanghai,
200234, P. R. China} \affiliation{$^{3}$Center of Theoretical
Nuclear Physics, National Laboratory of Heavy Ion Accelerator,
Lanzhou 730000, P. R. China} \affiliation{$^{4}$Institute of Modern
Physics, Lanzhou, 730000, P. R. China}

\begin{abstract}
We derive the generalized Friedmann equation governing the cosmological
evolution inside the thick brane model in the presence of two curvature
correction terms: a four-dimensional scalar curvature from induced gravity
on the brane, and a five-dimensional Gauss-Bonnet curvature term. We find
two effective four-dimensional reductions of the generalized Friedmann
equation in some limits and demonstrate that the reductions but not the
generalized Friedmann equation can be rewritten as the first law of
equilibrium thermodynamics on the apparent horizon of thick braneworld.
\end{abstract}

\maketitle

\section{Introduction}

Inspired by black hole thermodynamics, a profound connection between gravity
and thermodynamics has been argued to exist. In \cite{Jacobson} Jacobson
first showed that the Einstein gravity can be derived from the first law of
thermodynamics in the Rindler spacetime. For a general static spherically
symmetric spacetime, Padmanabhan pointed out that Einstein equations at the
horizon give rise to the first law of thermodynamics \cite{Padmanabhan}.
Recently the study on the connection between gravity and thermodynamics has
been extended to cosmological context. Frolov and Kofman in \cite{Frolov}
employed the approach proposed by Jacobson \cite{Jacobson} to a quasi-de
Sitter geometry of inflationary universe, and they calculated the energy
flux of a background slow-roll scalar through the quasi-de Sitter apparent
horizon. By applying the first law of thermodynamics to a cosmological
horizon, Danielsson obtained the Friedmann equation in the expanding
universe \cite{Danielsson}. In the quintessence dominated accelerating
universe, Bousso \cite{Bousso} showed that the first law of thermodynamics
holds at the apparent horizon. Cai and Kim \cite{Cai} generalized the
derivation of the Friedmann equations from the first law of thermodynamics
to the spacetime with any spatial curvature. This study has also been
generalized to the $f(R)$ gravity \cite{Eling,Akbar} and scalar-tensor
gravity theory \cite{Cao}. It has been disclosed that the first law of
thermodynamics can not be constructed unless introducing the non-equilibrium
entropy production term.

Besides gravity theories in four dimensions, the study on the connection
between gravity and thermodynamics has also been extended to the braneworld
cosmology \cite{Cao1,Sheykhi,Sheykhi1}. The main merit of braneworld
scenario is that it provided a novel approach to resolve the cosmological
constant and the hierarchy problems \cite{ADD,Randall}. It has been found
that the first law of thermodynamics on the apparent horizon can be derived
from the Friedmann equations in the Randall-Sundrum (RS) braneworld \cite%
{Cao1,Sheykhi}, and also in braneworld with curvature corrections including
the five-dimensional (5D) Gauss-Bonnet (GB) curvature correction \cite%
{Sheykhi1} and the four-dimensional (4D) scalar curvature from induced
gravity on the brane \cite{Sheykhi}. The former correction is inspired by
superstring theory which suggests the GB curvature correction as the first
and dominant quantum corrections to the Einstein-Hilbert action for a
ghost-free theory \cite{Zwiebach}. The combined action in five dimensions
gives the most general action with second-order field equation, as shown by
Lovelock \cite{Lovelock}. Their cosmological effects have been discussed in
several papers \cite{Nojiri}. The Dvali-Gabadadze-Porati (DGP) model
suggests the second curvature correction term to RS model, the 4D scalar
curvature term. This induced gravity correction term can be interpreted as
arising from a quantum effect due to the interaction between the bulk
gravitons and the matter on the brane \cite{Dvali}. The DGP scenario with GB
correction was also given, where the well-known DGP feature of late-time
acceleration without dark energy is preserved, but there is an intriguing
and new feature that the singularity in the early universe may be removed
\cite{Brown,Kofinas1}. In these braneworld scenarios the exact black hole
solution has not been found until now, and it was pointed out that the
connection between gravity and thermodynamics can shed lights on the entropy
of the braneworld \cite{Sheykhi,Sheykhi1}.

Although so many gravity theories have been linked to thermodynamics, it is
still unclear whether the connection always exists in arbitrary gravity
theories. Validating the connection in a more general gravity theory may be
helpful to answer the question. On the other hand, it is worth to note that
the previous mentioned braneworld scenarios are restricted by a simplifying
assumption that the brane is infinitely thin along the extra dimension with
solitonic localized matter distributions, and in more realistic models the
thickness of the brane should be taken into account. Study of thick branes
in the string inspired context of cosmology began almost simultaneously with
the study of thin brane. Different approaches were used to define and handle
the thickness. Most papers described the brane as the domain wall, based on
gravity coupled to scalars fields \cite{DeWolfe}. A smoothing or smearing
mechanism was used in \cite{Csaki} to modify the RS ansatz. Authors in \cite%
{Ghoroku} introduced a thickness to the brane by smoothing out the warp
factor of a thin braneworld to investigate the stability of a thick brane.
Mounaix and Langlois proposed a general approach to derive generalized
Friedmann equation, where the 4D effective brane quantities are obtained by
integrating the corresponding 5D ones along the extra dimension over the
brane thickness \cite{Mounaix}. In the case of a RS type cosmology, these
quantities are reduced to yield the cosmological equations in the low energy
limit and in the limit for a small brane thickness. When the brane thickness
is not small enough, it was disclosed that no effective 4D reduction is
possible unless some auxiliary quantities are introduced. Other methods have
also been presented recently, based on gluing a thick brane, considered as a
regular manifold, to two different manifolds on both sides of it \cite%
{Ghassemi}, and integrating the 5D Einstein equations along the fifth
dimension, while neglecting the parallel derivatives of the metric in
comparison with the transverse ones \cite{Navarro}. It is interesting to
note that the thickness effect has been considered to mimic the dark energy
by recent observations \cite{Movahed}.

In this paper, following Mounaix and Langlois's approach to define the 4D
effective brane quantities, we will study whether the thermodynamics may
relate to gravity on thick brane. To make our results with more generality,
we will not restrict us on the RS thick brane, instead we will generalize
the thick brane model in the presence of two curvature corrections: the 5D
GB curvature term and the 4D scalar curvature from induced gravity on the
brane. Compared with the investigation on the thermodynamics in other
gravity theories \cite{Cai, Cao, Cao1,Sheykhi,Sheykhi1}, it should be
stressed that this task is highly nontrival. This is partially because we
must obtain the exact Friedmann equations on thick brane with curvature
corrections and partially because the 4D effective thermodynamical brane
quantities have not been defined until now. We will derive the generalized
cosmological equations and obtain their 4D effective reductions. So one can
treat this work partially as the direct generalization of Mounaix and
Langlois's one, which will give access to a description of the intermediate
cases between the attractive DGP brane world with GB curvature correction
(where brane thickness is infinitely thin) and the opposite limit of an
infinitely thick brane, which effectively corresponds to the familiar 5D GB
gravity (since the infinitely thick 4D induced scalar curvature may be
absorbed into 5D Einstein-Hilbert scalar curvature). Furthermore, since our
work will make the braneworld scenarios discussed on \cite%
{Cao1,Sheykhi,Sheykhi1} more realistic and the gravity theory more general,
we can expect, besides helping the investigation on black hole of more
realistic braneworlds, to give further understanding on the deep connection
between gravity and thermodynamics.

This paper is arranged as follows. In section 2, we will generalize the
Friedmann equation of thick brane model with the 5D GB interaction and 4D
scalar curvature term. In the following section, we will reduce the
Friedmann equation in some limits. Then we will study the thermodynamical
behavior of the reduced Friedmann equations in section 4. Our conclusion and
discussion will be present in the last section.

\section{Cosmological equations for thick braneworlds}

Let us consider a thick braneworld model. For convenience and without loss
of generality we choose the extra dimension coordinates $y$ such that the
brane is center at $y=0$ and localized between $y=-1/2$ and $y=1/2$. For
simplicity, the spacetime has been restricted with $Z_{2}$ symmetry under
the transformation $y\rightarrow -y$. It is then always possible to find a
Gaussian coordinate system, starting from the hypersurface representing the
center\ of the brane. We consider an action which incorporates the induced
gravity and GB corrections%
\begin{equation}
S=\frac{1}{2\kappa _{5}^{2}}\int d^{5}x\sqrt{-g}\{\mathcal{L}_{EH}+\alpha
\mathcal{L}_{GB}\}+\frac{1}{2\kappa _{4}^{2}}\int_{-1/2}^{1/2}dy\int d^{4}x%
\sqrt{-\tilde{g}}\mathcal{L}_{IG},  \label{Sgrav}
\end{equation}%
where $\kappa _{5}$ ($\kappa _{4}$) is the bulk (brane) gravitational
constant and $g$ denotes the 5-dimensional bulk metric. The brane metric $%
\tilde{g}$ is defined as follows. For each fixed $y=y_{f}$ between $y=-1/2$
and $y=1/2$, the induced metric $\tilde{g}_{AB}\equiv g_{AB}-n_{A}n_{B}$,
where vector $n_{A}$ is normal to the slice of the brane at fixed $y_{f}$. $%
\mathcal{L}_{EH}=R-2\Lambda $ is the 5D Einstein-Hilbert Lagrangian with
negative cosmological constant $\Lambda <0$. The GB curvature correction
term $\mathcal{L}_{GB}$ is
\begin{equation*}
\mathcal{L}_{GB}=R^{2}-4R_{AB}R^{AB}+R_{ABCD}R^{ABCD}.
\end{equation*}%
We can define the GB coupling $\alpha $ through string energy scale $g_{s}$
as $\alpha =\frac{1}{8g_{s}^{2}}$. The second term of the action is the
generalization of the induced gravity action describing thin brane $\frac{1}{%
2\kappa _{4}^{2}}\int \delta (y)dy\int d^{4}x\sqrt{-\tilde{g}}\mathcal{L}%
_{IG}$ \cite{Dvali}. The induced gravity Lagrangian $\mathcal{L}_{IG}=\tilde{%
R}-2\kappa _{4}^{2}\lambda $ consists of 4D scale curvature $\tilde{R}$ and
brane tension $\lambda >0$, noting $\lambda $ is assumed as strictly
constant in $y$. We can define the crossover length scale of induced gravity
by $r_{c}=\kappa _{5}^{2}/\kappa _{4}^{2}$. For convenience, we will choose
the unit $\kappa _{5}=1$ throughout this paper. One can recover the thick RS
model when $r_{c}=\alpha =0$. The thick RS model with GB correction and the
thick DGP model correspond to the case with $r_{c}=0$ and $\alpha =0$,
respectively.

By varying the action in Eq. (\ref{Sgrav}) with respect to the bulk metric,
we obtain the field equation%
\begin{equation}
G_{AB}+2\alpha H_{AB}=\left. T_{AB}\right\vert _{total},
\label{field equation}
\end{equation}%
where $H_{AB}$ is the second order Lovelock tensor \cite{Lovelock}%
\begin{equation*}
H_{AB}=RR_{AB}-2R_{A}^{C}R_{BC}-2R^{CD}R_{ABCD}+R_{A}^{CDE}R_{BCDE}-\frac{1}{%
4}g_{AB}\mathcal{L}_{GB}.
\end{equation*}%
The total energy-momentum tensor $\left. T_{AB}\right\vert _{total}$ is
decomposed into bulk and brane components%
\begin{equation}
\left. T_{AB}\right\vert _{total}=\left. T_{AB}\right\vert _{bulk}+\left.
T_{AB}\right\vert _{brane}.  \label{et1}
\end{equation}%
The bulk component is
\begin{equation}
\left. T_{AB}\right\vert _{bulk}=-\Lambda g_{AB}.  \label{et2}
\end{equation}%
Since we are interested in the cosmological behavior of the braneworld we
take the metric ansatz%
\begin{equation}
ds^{2}=-n^{2}(t,y)dt^{2}+a^{2}(t,y)\gamma _{ij}dx^{i}dx^{j}+r_{b}^{2}dy^{2}
\label{metric}
\end{equation}%
where $\gamma _{ij}$ is a three-dimensional maximally symmetric metric whose
spatial curvature is characterized by $k=0,\pm 1$. For simplicity, we
consider the flat space $k=0$ in present work. The energy-momentum tensor of
the matter content in the brane is of the form
\begin{equation}
\left. T_{A}^{B}\right\vert _{brane}=\text{diag}\left[ \sqrt{\frac{\tilde{g}%
}{g}}(-\rho ,p,p,p),p_{T}\right] -\sqrt{\frac{\tilde{g}}{g}}r_{c}\tilde{G}%
_{A}^{B},  \label{et3}
\end{equation}%
where the brane tensor $\lambda $ is associated in%
\begin{equation}
\rho \equiv \rho _{m}+\lambda ,  \label{rolambda}
\end{equation}%
$\tilde{G}_{A}^{B}$ arises from the scalar curvature in Eq. (\ref{Sgrav}), $%
\rho $, $p$, $p_{T}$ are functions of $t$ and $y$, and $\sqrt{\frac{g}{%
\tilde{g}}}=r_{b}$ is the brane thickness which is assumed to be time
independent.

Our goal is to establish effective cosmological equations and study the
corresponding thermodynamics for an observer living in the brane. Because of
the finite thickness of the brane, there is some arbitrariness in the
definition of what the effective 4D quantities should be. Here we adopt the
prescription proposed by Mounaix and Langlois, in defining the 4D effective
quantity $\left\vert Q\right\vert (t)$ associated to a 5D quantity $Q(t,y)$
as its spatial average over the brane thickness \cite{Mounaix}
\begin{equation*}
\left\vert Q\right\vert (t)\equiv\left\langle Q(t,y)\right\rangle ,
\end{equation*}
where $\left\langle Q(t,y)\right\rangle =\int_{-1/2}^{1/2}Q(t,y)dy$. The 4D
"observable" counterparts of $a$, $\rho$ and $p$ thus are%
\begin{equation*}
\left\vert a\right\vert \equiv\left\langle a\right\rangle ,\;\left\vert
\rho\right\vert \equiv\left\langle \rho\right\rangle ,\;\left\vert
p\right\vert \equiv\left\langle p\right\rangle ,
\end{equation*}
and the observable Hubble parameter is%
\begin{equation}
\left\vert H\right\vert \equiv\frac{\left\langle \dot{a}\right\rangle }{%
\left\langle a\right\rangle }=H_{\left\langle a\right\rangle }.  \label{H}
\end{equation}
These observables are enough to establish effective cosmological equations.
Later we will further give the 4D observable counterparts about
thermodynamics.

Now, we study the components of field equations (\ref{field equation}), with
the metric (\ref{metric}) and energy-momentum tensors (\ref{et1}), (\ref{et2}%
) and (\ref{et3}). For later use, we define%
\begin{equation}
\Phi=\frac{\dot{a}^{2}}{n^{2}a^{2}}-\frac{a^{\prime2}}{r_{b}^{2}a^{2}},
\label{Fai0}
\end{equation}
where prime and dot denote the derivative with respect to $y$ and $t$,
respectively. Read $05$ component of the field equations (\ref{field
equation})%
\begin{equation*}
3(\frac{n^{\prime}}{n}\frac{\dot{a}}{a}-\frac{\dot{a}^{\prime}}{a}%
)(1+4\alpha\Phi)=0
\end{equation*}
which yields%
\begin{equation*}
n(t,y)=\xi(t)\dot{a}(t,y),
\end{equation*}
where $\xi$ depends on the normalization prescription for $n$. In following,
we will take the normalization $\left\langle n\right\rangle =1$ which gives%
\begin{equation}
\xi=\left\langle \dot{a}\right\rangle ^{-1}.  \label{05}
\end{equation}
The $00$ component of the field equations (\ref{field equation}) then reads%
\begin{equation}
(1+4\alpha\Phi)\frac{a^{\prime\prime}}{ar_{b}^{2}}=\Phi-\frac{1}{3}\left(
\frac{\rho}{r_{b}}-\frac{3r_{c}}{r_{b}}\frac{\dot{a}^{2}}{n^{2}a^{2}}%
+\Lambda\right) .  \label{du1}
\end{equation}
After inserting $\Phi$ (\ref{Fai0}), Eq. (\ref{du1}) can be rewritten as
following two useful expressions, under the normalization (\ref{05}),
\begin{equation}
(1+4\alpha\frac{\left\langle \dot{a}\right\rangle ^{2}}{a^{2}})\frac{\left(
a^{2}\right) ^{\prime\prime}}{2}-\frac{4\alpha}{r_{b}^{2}}a^{\prime2}\frac{%
a^{\prime\prime}}{a}-4\alpha\frac{\left\langle \dot{a}\right\rangle ^{2}}{%
a^{2}}a^{\prime2}=\left( r_{b}^{2}+r_{b}r_{c}\right) \left\langle \dot{a}%
\right\rangle ^{2}-\frac{1}{3}(r_{b}\rho a^{2}+r_{b}^{2}\Lambda a^{2}),
\label{junction}
\end{equation}
and%
\begin{equation}
\left[ (1+4\alpha\frac{\left\langle \dot{a}\right\rangle ^{2}}{a^{2}})\frac{%
\left( a^{2}\right) ^{\prime}}{2}\right] ^{\prime}-\frac{4\alpha }{3r_{b}^{2}%
}\left[ a^{2}\left( \frac{a^{\prime}}{a}\right) ^{3}\right] ^{\prime}+4\alpha%
\frac{\left\langle \dot{a}\right\rangle ^{2}}{a^{2}}a^{\prime2}-\frac{4\alpha%
}{3r_{b}^{2}}a^{2}\left( \frac{a^{\prime}}{a}\right) ^{4}=\left(
r_{b}^{2}+r_{b}r_{c}\right) \left\langle \dot {a}\right\rangle ^{2}-\frac{1}{%
3}(r_{b}\rho a^{2}+r_{b}^{2}\Lambda a^{2}).  \label{junction1}
\end{equation}
To deal with the complexity contributed by GB effect, we impose the condition%
\begin{equation}
a^{\prime2}\ll a\left\Vert a^{\prime\prime}\right\Vert ,  \label{thin}
\end{equation}
where $\left\Vert a^{\prime\prime}\right\Vert $ denotes the absolute value
of $a^{\prime\prime}$. It is known that, in the limit of thin brane, the
profiles for $a^{\prime\prime}$ blow up as delta function \cite{Binetruy},
hence we expect that the imposed condition (\ref{thin}) could be valid, at
least, in the brane with small (but finite) thickness. Thus the third term
on l.h.s. in Eq. (\ref{junction}) and the third and fourth terms on the
l.h.s. in Eq. (\ref{junction1}) can be neglected, which results in%
\begin{equation}
(1+4\alpha\frac{\left\langle \dot{a}\right\rangle ^{2}}{a^{2}})\frac{\left(
a^{2}\right) ^{\prime\prime}}{2}-\frac{4\alpha}{r_{b}^{2}}a^{\prime2}\frac{%
a^{\prime\prime}}{a}=\left( r_{b}^{2}+r_{b}r_{c}\right) \left\langle \dot{a}%
\right\rangle ^{2}-\frac{1}{3}(r_{b}\rho a^{2}+r_{b}^{2}\Lambda a^{2}),
\label{du2}
\end{equation}
and%
\begin{equation}
\left[ (1+4\alpha\frac{\left\langle \dot{a}\right\rangle ^{2}}{a^{2}})\frac{%
\left( a^{2}\right) ^{\prime}}{2}\right] ^{\prime}-\frac{4\alpha }{3r_{b}^{2}%
}\left[ a^{2}\left( \frac{a^{\prime}}{a}\right) ^{3}\right] ^{\prime}=\left(
r_{b}^{2}+r_{b}r_{c}\right) \left\langle \dot {a}\right\rangle ^{2}-\frac{1}{%
3}(r_{b}\rho a^{2}+r_{b}^{2}\Lambda a^{2}).  \label{du21}
\end{equation}
When GB effect disappears, the above two equations can also be obtained even
if the condition (\ref{thin}) is released.

Under $Z_{2}$ symmetry, integrating Eq. (\ref{du21}) over the brane, one can
obtain the boundary condition%
\begin{equation}
\left[ 1+\frac{8\alpha}{3}\frac{\left\langle \dot{a}\right\rangle ^{2}}{%
\left. a\right\vert _{\frac{1}{2}}^{2}}+\frac{4\alpha}{3}\left.
\Phi\right\vert _{\frac{1}{2}}\right] \left. \left( aa^{\prime}\right)
\right\vert _{\frac{1}{2}}=\frac{1}{2}\left( r_{b}^{2}+r_{b}r_{c}\right)
\left\langle \dot{a}\right\rangle ^{2}-\frac{1}{6}(r_{b}\left\langle \rho
a^{2}\right\rangle +r_{b}^{2}\Lambda\left\langle a^{2}\right\rangle ).
\label{bc2}
\end{equation}
Solving $\left. a^{\prime}\right\vert _{\frac{1}{2}}$ from that, we have%
\begin{equation}
\left. a^{\prime}\right\vert _{\frac{1}{2}}=\frac{1}{\gamma}\frac {%
\left\langle a\right\rangle }{\beta}\left( \frac{1}{2}\zeta
r_{b}^{2}H_{\left\langle a\right\rangle }^{2}-\varepsilon\eta-\frac{1}{6}%
r_{b}^{2}\Lambda\tilde{\eta}\right) ,  \label{ap}
\end{equation}
with some dimensionless quantities have been used:%
\begin{equation*}
\varepsilon\equiv\frac{1}{6}r_{b}\left\langle \rho\right\rangle
\end{equation*}%
\begin{equation*}
\gamma\equiv1+\frac{8\alpha}{3\beta^{2}}H_{\left\langle a\right\rangle }^{2}+%
\frac{4\alpha}{3}\left. \Phi\right\vert _{\frac{1}{2}}
\end{equation*}%
\begin{equation*}
\eta\equiv\frac{\left\langle \rho a^{2}\right\rangle }{\left\langle
\rho\right\rangle \left\langle a\right\rangle ^{2}},\;\tilde{\eta}\equiv
\frac{\left\langle a^{2}\right\rangle }{\left\langle a\right\rangle ^{2}}%
,\;\beta\equiv\frac{\left. a\right\vert _{1/2}}{\left\langle a\right\rangle }%
,\;\zeta=\left( 1+\frac{r_{c}}{r_{b}}\right) .
\end{equation*}
Whereas $\varepsilon$ characterizes the thickness of the brane, the
quantities $\beta,\eta,\tilde{\eta}$ characterize the inhomogeneity of the
brane along the fifth dimension (in the case of a homogeneous brane, one has
$\beta =\eta=\tilde{\eta}=1$). The quantities $\gamma$ and$\;\zeta$ embody
the GB effect and induced gravity effect, respectively (in the case of RS
thick brane, one has $\gamma=\zeta=1$).

The $55$ component of the field equation (\ref{field equation}) can be
rewritten as%
\begin{equation*}
\dot{F}=\frac{2}{3}a^{3}\dot{a}P_{T}
\end{equation*}
where
\begin{equation*}
F=a^{4}(\Phi+2\alpha\Phi^{2}-\frac{1}{6}\Lambda).
\end{equation*}
Imposing the boundary condition $P_{T}(\pm\frac{1}{2})=0$, one has $\dot {F}%
(\pm\frac{1}{2})=0$, which gives, after time integration,%
\begin{equation}
r_{b}^{2}H_{\left\langle a\right\rangle }^{2}=\frac{\left. a^{\prime
2}\right\vert _{1/2}}{\left\langle a\right\rangle ^{2}}+\left. \Phi
\right\vert _{\frac{1}{2}}r_{b}^{2}\beta^{2},  \label{jc2}
\end{equation}
where%
\begin{equation}
\left. \Phi\right\vert _{\frac{1}{2}}=\frac{-1\pm\sqrt{1+\frac{4}{3}%
\alpha\left( \Lambda+\frac{C}{\beta^{2}\left\langle a\right\rangle ^{4}}%
\right) }}{4\alpha}.  \label{Fai}
\end{equation}
Hereafter, we will select the above branch of (\ref{Fai}) for correct limit
in $\alpha\rightarrow0$,%
\begin{equation*}
\left. \Phi\right\vert _{\frac{1}{2}}=\frac{1}{6}\Lambda,
\end{equation*}
and omit the dark radiation term $C=0$, since in this paper we only care of
AdS$_{5}$ bulk. The AdS$_{5}$ length scale associated to the (negative)
cosmological constant in the bulk is defined by%
\begin{equation*}
l_{\Lambda}\equiv\sqrt{-\frac{6}{\Lambda}}.
\end{equation*}
Substituting $\left. a^{\prime2}\right\vert _{1/2}$ (\ref{ap}) into Eq. (\ref%
{jc2}), one has%
\begin{equation*}
\beta^{2}\gamma^{2}r_{b}^{2}H_{\left\langle a\right\rangle }^{2}=\left(
\frac{1}{2}\zeta r_{b}^{2}H_{\left\langle a\right\rangle }^{2}-\varepsilon
\eta+\frac{r_{b}^{2}}{l_{\Lambda}^{2}}\tilde{\eta}\right) ^{2}+\left.
\Phi\right\vert _{\frac{1}{2}}r_{b}^{2}\beta^{4}\gamma^{2}.
\end{equation*}
Then the generalized Friedmann equation can be obtained%
\begin{equation}
H_{\left\langle a\right\rangle }^{2}=\frac{2}{r_{b}^{2}}(\frac{%
\beta^{2}\gamma^{2}}{\zeta^{2}}+\frac{\varepsilon\eta}{\zeta}-\frac{r_{b}^{2}%
\tilde{\eta}}{l_{\Lambda}^{2}\zeta})\left[ 1\pm\sqrt{1-\frac{\left(
\varepsilon\eta-\tilde{\eta}r_{b}^{2}/l_{\Lambda}^{2}\right) ^{2}+\gamma
^{2}\left. \Phi\right\vert _{\frac{1}{2}}r_{b}^{2}\beta^{4}}{\zeta^{2}(\frac{%
\beta^{2}\gamma^{2}}{\zeta^{2}}+\frac{\varepsilon\eta}{\zeta}-\frac{r_{b}^{2}%
\tilde{\eta}}{l_{\Lambda}^{2}\zeta})^{2}}}\right] .  \label{general FM}
\end{equation}
Hereafter, we will select the below branch for correct thin brane limit in $%
r_{b}\rightarrow0$. This is the generalized Friedmann equation governing the
cosmological evolution inside the thick brane, in the presence of two
curvature correction terms. It is important to notice that this equation
should be regraded as an implicit equation for $H_{\left\langle
a\right\rangle }$ because $\beta,\eta,\tilde{\eta},\gamma$ can depend on $%
H_{\left\langle a\right\rangle }$.

It is instructive to consider two opposite limits of small and large $r_{b}$%
. In the limit $r_{b}\rightarrow 0$, Eq. (\ref{general FM}) reduces to%
\begin{equation*}
H_{\left\langle a\right\rangle }^{2}=\frac{2\beta ^{2}\gamma ^{2}}{r_{c}^{2}}%
+\frac{\left\langle \rho \right\rangle \eta }{3r_{c}}-\frac{1}{\sqrt{3}r_{c}}%
\sqrt{\frac{12\beta ^{4}\gamma ^{4}}{r_{c}^{2}}+\frac{4\beta ^{2}\gamma
^{2}\left\langle \rho \right\rangle \eta }{r_{c}}-12\beta ^{4}\gamma
^{2}\left. \Phi \right\vert _{\frac{1}{2}}},
\end{equation*}%
where the inhomogeneity parameters $\beta ,\eta ,\tilde{\eta}$ tend to $1$
in this limit. One can find that it exactly recovers the junction condition
of thin brane cosmology with curvature corrections \cite{Kofinas1}%
\begin{equation}
4\gamma ^{2}\left( H_{\left\langle a\right\rangle }^{2}-\left. \Phi
\right\vert _{\frac{1}{2}}\right) =\left( r_{c}H_{\left\langle
a\right\rangle }^{2}-\frac{\left\langle \rho \right\rangle }{3}\right) ^{2}.
\label{jun}
\end{equation}%
It is a cubic equation of $H_{\left\langle a\right\rangle }^{2}$ and we do
not write its expatiatory solution clearly (one can find the exact solution
in \cite{Kofinas1}). When GB effect disappears, we can discuss the opposite
limit $r_{b}\rightarrow \infty $ and simplify the Eq. (\ref{general FM}) as%
\begin{equation*}
H_{\left\langle a\right\rangle }^{2}=\frac{\left\langle \rho \right\rangle
\eta }{3r_{b}}-\frac{2\tilde{\eta}}{l_{\Lambda }^{2}}+\frac{2r_{c}}{%
r_{b}l_{\Lambda }^{2}}+\frac{2\beta }{r_{b}}\sqrt{\frac{-2\tilde{\eta}^{2}}{%
l_{\Lambda }^{2}}+\frac{\beta ^{2}}{l_{\Lambda }^{2}}},
\end{equation*}%
where terms to $O(r_{b}^{-1})$ have been kept, assuming that the
inhomogeneity parameters remain bounded as $r_{b}\rightarrow \infty $. In
the case of a Minkowski bulk, $\Lambda =0$, one obtains%
\begin{equation}
H_{\left\langle a\right\rangle }^{2}=\frac{\left\langle \rho \right\rangle
\eta }{3r_{b}}  \label{jun1}
\end{equation}%
which corresponds to the standard Friedmann equation. It is interesting to
note that the induced gravity correction does not affect the Friedmann
equation and the limit $r_{b}\rightarrow \infty $ allows the possibility of
a homogeneous brane along the fifth dimension $\eta =1$, in agreement with
the usual Kaluza-Klein picture.

At the end of this section, we now consider the conservation of matter on
thick brane. Since the l.h.s in the field equation (\ref{field equation}) is
divergence-free, the total energy-momentum tensor is conserved in the bulk%
\begin{equation}
\nabla_{A}\left. T_{B}^{A}\right\vert _{total}=0.  \label{sss}
\end{equation}
Using the FRW metric (\ref{metric}), the zero component of Eq. (\ref{sss})
yields%
\begin{equation}
\dot{\rho}+3H(\rho+p)=0.  \label{ro}
\end{equation}

\section{The effective 4D reductions of cosmological equations}

There are two effective 4D reductions of cosmological equations in RS thick
brane, in the low energy limit and in the limit for a small brane thickness
\cite{Mounaix}. Accordingly, we will reduce the Friedmann equation (\ref%
{general FM}) and the conservation equation (\ref{ro}).

\subsection{Cosmological equations in the low energy limit}

In the thin brane cosmology, the brane tensor is adjusted to compensate the
bulk cosmological constant, corresponding to the fine-turning condition of
RS model, which can be reexpressed as a relation%
\begin{equation}
l_{\lambda}=l_{\Lambda},  \label{RS fine turn}
\end{equation}
where the length scale is defined from the brane tension $\lambda$ as $%
l_{\lambda}\equiv6/\lambda$. In the current theory, we generalize the
version proposed in \cite{Mounaix} for RS thick brane%
\begin{equation*}
l_{\lambda}\equiv\frac{\tilde{\eta}(1-u)}{\beta^{2}}l_{\Lambda}
\end{equation*}
to thick brane with GB curvature correction (while the induced gravity
correction does not affect the fine-turning relation)%
\begin{equation}
l_{\lambda}\equiv\frac{\tilde{\eta}(1-u)}{\beta^{2}\left( 1+\frac{4}{3}%
\alpha\left. \Phi\right\vert _{\frac{1}{2}}\right) }\tilde{l}_{\Lambda },
\label{llamda}
\end{equation}
where the dimensionless ratio has been introduced%
\begin{equation}
u\equiv\frac{r_{b}l_{\lambda}}{l_{\Lambda}^{2}}=-\frac{r_{b}\Lambda}{\lambda
},  \label{u}
\end{equation}
and%
\begin{equation*}
\tilde{l}_{\Lambda}\equiv\left( -\left. \Phi\right\vert _{\frac{1}{2}%
}\right) ^{-\frac{1}{2}}.
\end{equation*}
can be referred as corrected AdS$_{5}$ length scale. In the limit where $%
r_{b}$ and $\alpha$ go to zero, it can be checked that Eq. (\ref{llamda})
reduces to RS condition (\ref{RS fine turn}), as it should be.

Now we will consider the low energy regime in which the ordinary energy (and
pressure) is small with respect to the brane tensor. To do that, we
introduce the dimensionless parameters%
\begin{equation*}
\varepsilon_{\lambda}\equiv\frac{r_{b}}{l_{\lambda}},\;\varepsilon_{\rho
}\equiv\frac{\eta\left\langle \rho_{m}\right\rangle }{\lambda},
\end{equation*}
where%
\begin{equation*}
\eta\equiv\frac{\left\langle \rho_{m}a^{2}\right\rangle }{\left\langle
\rho_{m}\right\rangle \left\langle a^{2}\right\rangle }.
\end{equation*}
Considering the low energy limit in the regime defined by $\varepsilon_{\rho
}\ll\min(1,1/\varepsilon_{\lambda})$, and assuming the GB effect is first
order small in $\varepsilon_{\rho}$, one can find that at lowest order in $%
\varepsilon_{\rho}$, Eq. (\ref{general FM}) can be expanded as
\begin{equation*}
\zeta r_{b}^{2}H_{\left\langle a\right\rangle }^{2}=\frac{2\varepsilon
_{\lambda}^{2}\varepsilon_{\rho}}{\varepsilon_{\lambda}+\frac{\tilde {l}%
_{\Lambda}}{\zeta l_{\lambda}}}-\frac{8\alpha\varepsilon_{\lambda}^{2}%
\varepsilon_{\rho}}{\zeta^{2}\tilde{l}_{\Lambda}l_{\lambda}^{3}\left(
\varepsilon_{\lambda}+\frac{\tilde{l}_{\Lambda}}{\zeta l_{\lambda}}\right)
^{2}},
\end{equation*}
and the quantities $\beta,\eta,\tilde{\eta}$ must be determined at zeroth
order in $\varepsilon_{\rho}$. To obtain an effective reduction, in the
limit at zeroth order in $\varepsilon_{\rho}$, the 00 component of Einstein
equations (see Eq. (\ref{du1}))%
\begin{equation*}
(1+4\alpha\Phi)\frac{a^{\prime\prime}}{ar_{b}^{2}}=\Phi-\frac{1}{3}\left[
\frac{\rho}{r_{b}}-\frac{3r_{c}}{r_{b}}\frac{\left\langle \dot{a}%
\right\rangle ^{2}}{a^{2}}+\Lambda\right]
\end{equation*}
reduces to%
\begin{align}
\left( \frac{a^{2}}{\left\langle a\right\rangle ^{2}}\right) ^{\prime\prime}
& =2\zeta r_{b}^{2}H_{\left\langle a\right\rangle }^{2}-4\varepsilon
_{\lambda}(1-u)\frac{a^{2}}{\left\langle a\right\rangle ^{2}}  \notag \\
& =-4\varepsilon_{\lambda}(1-u)\frac{a^{2}}{\left\langle a\right\rangle ^{2}}%
,  \label{a2DGP}
\end{align}
where we have neglected the terms $4\alpha\Phi$ and $2\zeta
r_{b}^{2}H_{\left\langle a\right\rangle }^{2}$ which are first order in $%
\varepsilon_{\rho}$. Restricting ourselves to the case of physical interest $%
(1-u)>0$, and integrating Eq. (\ref{a2DGP}) with the boundary condition $%
a^{\prime}(y=0)=0$, we have%
\begin{equation}
a=\left\langle a\right\rangle \left[ B\cos(ky)\right] ^{1/2},
\label{a low energy}
\end{equation}
with%
\begin{equation}
k^{2}=4\varepsilon_{\lambda}(1-u)  \label{k2}
\end{equation}
and%
\begin{equation}
B=\left\langle \cos^{1/2}(ky)\right\rangle ^{-2}.  \label{B}
\end{equation}
Simultaneously, one can also obtain%
\begin{equation*}
\beta=\left[ B\cos(k/2)\right] ^{1/2}
\end{equation*}%
\begin{equation}
\tilde{\eta}=2\frac{B}{k}\sin\left( k/2\right) .  \label{Etaba}
\end{equation}
Here we necessarily have $k\prec\pi$, and the scale factor $a$ is always
positive throughout the brane, reaching the extremity of the brane at $y=1/2$%
. Now we can determine the evolution of $\rho_{m}$ at lowest order in $%
\varepsilon_{\rho}$. Since neither $B$ nor $k$ depends on $t$, the
conservation equation (\ref{ro}) simply yields%
\begin{equation}
\dot{\rho}_{m}+3H_{\left\langle a\right\rangle }(\rho_{m}+p_{m})=0,
\label{ro low energy}
\end{equation}
from which it follows that, at this order, $\rho_{m}$ factorizes as $\rho
_{m}(y,t)=f(y)\left\langle \rho_{m}\right\rangle (t)$. Then one gets%
\begin{equation*}
\eta=\frac{\left\langle f(y)\cos(ky)\right\rangle }{\left\langle
f(y)\right\rangle \left\langle \cos^{1/2}(ky)\right\rangle ^{2}}.
\end{equation*}
The effective 4D thick brane cosmology with induced gravity correction and
the first order small GB effect in the limit $\varepsilon_{\rho}\ll\min
(1,1/\varepsilon_{\lambda})$ is thus given by%
\begin{equation}
H_{\left\langle a\right\rangle }^{2}=\frac{\eta}{3\left( \zeta r_{b}+\tilde{l%
}_{\Lambda}\right) }\left\langle \rho_{m}\right\rangle -\frac{4\alpha}{3}%
\frac{\eta}{\tilde{l}_{\Lambda}\left( \zeta r_{b}+\tilde {l}%
_{\Lambda}\right) ^{2}}\left\langle \rho_{m}\right\rangle ,
\label{low energy H2}
\end{equation}
and%
\begin{equation*}
\left\langle \dot{\rho}_{m}\right\rangle +3H_{\left\langle a\right\rangle
}(\left\langle \rho_{m}\right\rangle +\left\langle p_{m}\right\rangle )=0.
\end{equation*}
The thickness effect is embodied by $\eta$ besides $r_{b}$, which, in this
limit, are constant. The induced gravity effect is embodied by constant $%
\zeta$, and the GB effect is embodied by $\alpha$. Inserting the
fine-turning condition (\ref{llamda}) under infinitely thin limit $%
r_{b}\rightarrow0$ into the junction condition (\ref{jun}) for thin
braneworld with induced gravity correction and the first order small GB
effect at low energy, one can solve%
\begin{equation*}
H_{\left\langle a\right\rangle }^{2}=\frac{\lambda\left\langle \rho
_{m}\right\rangle }{3(6+3r_{c}\lambda)}-\frac{4\alpha\lambda^{3}\left\langle
\rho_{m}\right\rangle }{27(6+3r_{c}\lambda)^{2}},
\end{equation*}
which can be exactly recovered from the reduced Friedmann equation (\ref{low
energy H2}) under infinitely thin limit $r_{b}\rightarrow0$. Also, one can
easily find that Eq. (\ref{low energy H2}) recovers the reduction of thick
RS braneworld when $\zeta=1$ and $\alpha=0$ \cite{Mounaix}.

Let us discuss the condition (\ref{thin}) in current limit. Using
fine-tuning condition (\ref{llamda}) with Eqs. (\ref{k2}), (\ref{B}), and (%
\ref{Etaba}), we obtain%
\begin{equation*}
k=2\tan^{-1}\sqrt{\frac{-6r_{b}\left. \Phi\right\vert _{1/2}(1+\frac{4}{3}%
\alpha\left. \Phi\right\vert _{1/2})^{2}}{\lambda+r_{b}\Lambda}.}
\end{equation*}
The infinitely thin limit $r_{b}\rightarrow0$ corresponds to $k\rightarrow0$%
. Using Eq. (\ref{a low energy}), one can find that the condition (\ref{thin}%
) now takes the form%
\begin{equation*}
\frac{2\sin^{2}(ky)}{3+\cos(2ky)}\ll1,
\end{equation*}
which is naturally satisfied for small but finite brane thickness.

At last, for our later thermodynamic use, we solve $\left\langle \rho
_{m}\right\rangle $ clearly from Eq. (\ref{low energy H2})
\begin{equation}
\left\langle \rho_{m}\right\rangle =\frac{H_{\left\langle a\right\rangle
}^{2}}{\eta}\left[ \frac{18}{\lambda}+3\left( r_{b}+r_{c}\right) +\frac{%
4\alpha\lambda}{3}\right] .  \label{ro final}
\end{equation}

\subsection{Cosmological equations for a small brane thickness}

Next we wish to discuss another limit case. In order to simplify the
calculations, we will restrict ourselves to the case where the bulk
cosmological constant vanishes, i.e. the space-time outside the brane
effectively is a 5D Minkowski space-time. In this case, one has $\Lambda
=\left. \Phi\right\vert _{\frac{1}{2}}=l_{\Lambda}^{-2}=\tilde{l}_{\Lambda
}^{-2}=0$ and Eq. (\ref{general FM}) reads%
\begin{equation}
H_{\left\langle a\right\rangle }^{2}=\frac{2}{\zeta r_{b}^{2}}(\frac{\beta
^{2}\gamma^{2}}{\zeta}+\varepsilon\eta)\left[ 1-\sqrt{1-\frac{\left(
\varepsilon\eta\right) ^{2}}{(\frac{\beta^{2}\gamma^{2}}{\zeta}%
+\varepsilon\eta)^{2}}}\right] .  \label{H3}
\end{equation}
In the following, we will consider the limit $\varepsilon\ll1$. In contrast
with previous subsection where the GB effect is assumed to be small, now we
assume the induced gravity effect is small $\zeta\lesssim\gamma^{2}$. At
third order in $\varepsilon$, Eq. (\ref{H3}) can be written as%
\begin{equation*}
\zeta r_{b}^{2}H_{\left\langle a\right\rangle }^{2}=\frac{\left(
\varepsilon\eta\right) ^{2}}{\frac{\beta^{2}\gamma^{2}}{\zeta}+\varepsilon
\eta}.
\end{equation*}
At this order, \bigskip$\beta$ and $\eta$ must be expressed at first and
zeroth order respectively. To determine $a$ perturbatively in $\varepsilon$
we will use the following $y$ expansions%
\begin{align*}
a(t,y) & =\left\langle a\right\rangle (t)\sum_{n=0}^{+\infty}\bar{a}%
_{n}(t)y^{2n}\equiv\left\langle a\right\rangle (t)\Sigma_{a}(t,y), \\
\rho(t,y) & =\left\langle \rho\right\rangle (t)\sum_{n=0}^{+\infty}\bar{\rho
}_{n}(t)y^{2n}\equiv\left\langle \rho\right\rangle (t)\Sigma_{\rho}(t,y).
\end{align*}
Inserting these expansions in Eq. (\ref{du2}) with $\Lambda=0$, one obtains%
\begin{equation}
(1+4\alpha H_{\left\langle a\right\rangle }^{2}\frac{1}{\Sigma_{a}^{2}}%
)\left( \Sigma_{a}^{2}\right) ^{\prime\prime}-\frac{8\alpha}{r_{b}^{2}}%
\left( \Sigma_{a}^{\prime}\right) ^{2}\frac{\Sigma_{a}^{\prime\prime}}{%
\Sigma_{a}}=2\zeta r_{b}^{2}H_{\left\langle a\right\rangle
}^{2}-4\varepsilon\Sigma_{\rho}\Sigma_{a}^{2}.  \label{sg2}
\end{equation}
Since one needs $a$ at first order in $\varepsilon$, one can neglect the
first term on r.h.s in Eq. (\ref{sg2}), and consider the constant (in $y$)
component of Eq. (\ref{sg2}) only. We therefore get%
\begin{equation*}
(1+4\alpha H_{\left\langle a\right\rangle }^{2}\frac{1}{\bar{a}_{0}^{2}})4%
\bar{a}_{0}\bar{a}_{1}=-4\varepsilon\bar{\rho}_{0}\bar{a}_{0}^{2}.
\end{equation*}
$\bar{a}_{1}$ can be solved from this equation%
\begin{equation}
\bar{a}_{1}=-\frac{\varepsilon}{\theta}\bar{\rho}_{0}\bar{a}_{0},  \label{a1}
\end{equation}
where the dimensionless parameter is%
\begin{equation*}
\theta=(1+4\alpha H_{\left\langle a\right\rangle }^{2}\frac{1}{\bar{a}%
_{0}^{2}}).
\end{equation*}
Thus, we have%
\begin{equation}
a=\left\langle a\right\rangle \bar{a}_{0}(1-\frac{\varepsilon}{\theta}\bar{%
\rho}_{0}y^{2}).  \label{aa}
\end{equation}
In this expression, $\theta$ and $\bar{\rho}_{0}$ must be determined at
zeroth order in $\varepsilon$, which yields $\bar{\rho}_{0}=1$ and $%
\left\langle \rho\right\rangle $ as the solution to the usual 4D
energy-momentum conservation equation. From Eq. (\ref{aa}) at first order in
$\varepsilon$, one has%
\begin{equation*}
\bar{a}_{0}=\frac{1}{1-\frac{1}{12}\frac{\varepsilon}{\theta}},
\end{equation*}
and%
\begin{equation*}
\theta=1+4\alpha H_{\left\langle a\right\rangle }^{2}(1-\frac{1}{6}\frac{%
\varepsilon}{\theta}).
\end{equation*}
Since $\theta$ must be determined at zeroth order in $\varepsilon$, one has%
\begin{equation}
\theta=1+4\alpha H_{\left\langle a\right\rangle }^{2}.  \label{Theta}
\end{equation}
Eq. (\ref{aa}) finally reads%
\begin{equation}
a=\left\langle a\right\rangle \left[ 1-\frac{\varepsilon}{\theta}(y^{2}-%
\frac{1}{12})\right] ,  \label{a small thickness}
\end{equation}
and the inhomogeneity parameters are $\beta=1-\frac{1}{6}\frac{\varepsilon }{%
\theta}$ and $\eta=1$. Using Eq. (\ref{a small thickness}), one can find
that the condition (\ref{thin}) is naturally satisfied since $%
\varepsilon\ll1 $.

Now, at third order in $\varepsilon$, the quantity $\zeta
r_{b}^{2}H_{\left\langle a\right\rangle }^{2}$ reads%
\begin{equation}
\zeta r_{b}^{2}H_{\left\langle a\right\rangle }^{2}=\frac{\varepsilon^{2}}{%
\frac{\left( 1-\frac{1}{3}\frac{\varepsilon}{\theta}\right) \gamma^{2}}{\zeta%
}+\varepsilon},  \label{thirdH3}
\end{equation}
or, equivalently,%
\begin{equation}
H_{\left\langle a\right\rangle }^{2}=\frac{\frac{\left\langle \rho
\right\rangle ^{2}}{36}}{\left( 1-\frac{1}{18}\frac{r_{b}\left\langle
\rho\right\rangle }{\theta}\right) \gamma^{2}+\frac{(r_{b}+r_{c})\left%
\langle \rho\right\rangle }{6}},  \label{ro4}
\end{equation}
where%
\begin{equation*}
\gamma=1+\frac{8\alpha}{3\left( 1-\frac{1}{18}\frac{r_{b}\left\langle
\rho\right\rangle }{\theta}\right) }H_{\left\langle a\right\rangle }^{2}.
\end{equation*}
The effective 4D Friedmann equation (\ref{thirdH3}) and conservation equation%
\begin{equation}
\left\langle \dot{\rho}\right\rangle +3H_{\left\langle a\right\rangle
}(\left\langle \rho\right\rangle +\left\langle p\right\rangle )=0
\label{ro small thickness}
\end{equation}
compose the effective reduced 4D cosmological equations on thick brane.

The equation of $\zeta r_{b}^{2}H_{\left\langle a\right\rangle }^{2}$ is
fourth order, whose exact solution is expatiatory and we do not give the
exact solution. But one can find that the junction condition (\ref{jun}) of
thin braneworld with GB correction and small induced gravity correction can
be recovered from Eq. (\ref{ro4}) under limit $r_{b}\rightarrow0$. An
interesting case is when $\alpha$ is very big $\alpha H_{\left\langle
a\right\rangle }^{2}>\varepsilon^{-1}$, Eq. (\ref{thirdH3}) can be
simplified as%
\begin{equation*}
\zeta r_{b}^{2}H_{\left\langle a\right\rangle }^{2}=\frac{\zeta\varepsilon
^{2}}{(\frac{8\alpha}{3}H_{\left\langle a\right\rangle }^{2})^{2}}
\end{equation*}
or, equivalently,%
\begin{equation*}
H_{\left\langle a\right\rangle }^{2}=\left( \frac{\left\langle \rho
\right\rangle }{4\alpha}\right) ^{\frac{2}{3}},
\end{equation*}
which tells us that as GB effect is big, the effect of small thickness of
brane (also small induced gravity effect) can be neglected. Moreover, for
later thermodynamic use, we write $\left\langle \rho\right\rangle $ from Eq.
(\ref{ro4}) clearly%
\begin{equation}
\left\langle \rho\right\rangle =\frac{1}{I}\left\{ 9\left[ H_{\left\langle
a\right\rangle }^{2}\theta J+\sqrt{4(3+8\alpha)^{2}H_{\left\langle
a\right\rangle }^{2}\theta^{2}I-H_{\left\langle a\right\rangle }^{2}J^{2}}%
\right] \right\} ,  \label{roH2 small thickness}
\end{equation}
with%
\begin{align*}
I & =81\theta^{2}+16\alpha H_{\left\langle a\right\rangle }^{2}\left(
3+8\alpha\right) r_{b}^{2}, \\
J & =39+16\alpha H_{\left\langle a\right\rangle }^{2}(5-4\alpha
H_{\left\langle a\right\rangle }^{2})+27\theta r_{c}.
\end{align*}
We are also interested in $\left\langle \rho\right\rangle $ at first order
of $r_{b}$, $r_{c}$ and $\alpha$%
\begin{equation}
\left\langle \rho\right\rangle =6H_{\left\langle a\right\rangle
}+3H_{\left\langle a\right\rangle }^{2}r_{c}+\frac{22}{3}H_{\left\langle
a\right\rangle }^{2}r_{b}+16\alpha H_{\left\langle a\right\rangle }^{3}-%
\frac{76}{9}\alpha H_{\left\langle a\right\rangle }^{4}r_{b}.
\label{roH2 simplify}
\end{equation}

\section{The first law of thermodynamics on apparent horizon of FRW thick
brane}

Now we try to construct the first law of thermodynamics on apparent horizon
of thick brane. To have further understanding about the nature of apparent
horizon we rewrite more explicitly, the 4D metric of FRW universe on the
brane in the form%
\begin{equation*}
ds^{2}=h_{ab}dx^{a}dx^{b}+\tilde{r}^{2}d\Omega _{2}^{2},
\end{equation*}%
where $h_{ab}=$diag$(-1,\frac{a^{2}}{1-ka^{2}})$, $d\Omega _{2}^{2}$ is the $%
2$-dimensional sphere element, and $x^{0}=t,\;x^{1}=r,\;\tilde{r}=ar$ is the
radius of the sphere and $a$ is the scale factor. For simplicity, we
consider the flat space $k=0$ in this paper. It is known that the dynamical
apparent horizon of thin brane, the marginally trapped surface with
vanishing expansion, is defined as a sphere situated at $r=r_{A}$ satisfying%
\begin{equation*}
h^{ab}\partial _{a}\tilde{r}\partial _{b}\tilde{r}=0,
\end{equation*}%
which can be solved explicitly,%
\begin{equation*}
r_{A}=\frac{1}{\sqrt{\frac{\dot{a}^{2}}{n^{2}}}}.
\end{equation*}%
The sphere has radius%
\begin{equation*}
\tilde{r}_{A}\equiv r_{A}a=\frac{1}{\sqrt{\frac{\dot{a}^{2}}{n^{2}a^{2}}}}.
\end{equation*}%
The associated temperature on the apparent horizon is%
\begin{equation}
T\equiv \frac{\kappa }{2\pi },  \label{T}
\end{equation}%
where $\kappa $ is the surface gravity%
\begin{equation*}
\kappa \equiv \left. \frac{1}{\sqrt{-h}}\partial _{a}\left( \sqrt{-h}%
h^{ab}\partial _{b}\tilde{r}\right) \right\vert _{r=r_{A}}.
\end{equation*}%
Define $T_{a}^{b}$ as the projection of the 4-dimensional energy-momentum
tensor $T_{\nu }^{\mu }$ of a perfect fluid matter in the FRW universe in
the normal direction of the 2-sphere. We have work density \cite{Hayward}%
\begin{equation*}
W\equiv -\frac{1}{2}T_{a}^{a},
\end{equation*}%
and energy-supply vector%
\begin{equation*}
\Psi _{a}\equiv T_{a}^{b}\partial _{b}\tilde{r}+W\partial _{a}\tilde{r}.
\end{equation*}%
Expressing $A=4\pi \tilde{r}_{A}^{2}$ and$\;V=\frac{4}{3}\pi \tilde{r}%
_{A}^{3}$ as the area and volume of an 3-dimensional space with radius $%
\tilde{r}_{A}$ respectively, we can write the total energy flux on the
apparent horizon as%
\begin{equation*}
\nabla E\equiv A\Psi +W\nabla V.
\end{equation*}%
Because of the finite thickness of the brane, there is some arbitrariness in
the definition of what the effective 4D quantities should be. Following the
order to define the radius of apparent horizon, its area and volume, the
temperature on it, the projection of the 4-dimensional energy-momentum
tensor, the work density done by a change of it and energy-supply vector
flow through it, and finally the total energy flux on it, we define their
effective 4D quantities as follows. The location of apparent horizon is
determined by%
\begin{equation*}
\left\langle h^{ab}\partial _{a}\tilde{r}\partial _{b}\tilde{r}\right\rangle
=0,
\end{equation*}%
which has the solution%
\begin{equation*}
\left\vert r_{A}\right\vert \equiv \frac{1}{\sqrt{\left\langle \frac{\dot{a}%
^{2}}{n^{2}}\right\rangle }}=\frac{1}{\left\langle \dot{a}\right\rangle }.
\end{equation*}%
The radius of horizon is set as {\footnote{%
Remembering that the effective reductions of Friedmann equation in previous
section need the quantity $\zeta r_{b}^{2}H_{\left\langle a\right\rangle
}^{2}$ is very small, here we like to note that the reductions are carried
out when the horizon is very big than brane thickness and crossover scale.}%
\begin{equation}
\left\vert \tilde{r}_{A}\right\vert \equiv \left\langle \left\vert
r_{A}\right\vert a\right\rangle =\frac{1}{H_{\left\langle a\right\rangle }}.
\label{horizon}
\end{equation}%
The temperature can be obtained through surface gravity%
\begin{equation}
\left\vert T\right\vert \equiv \frac{\left\vert \kappa \right\vert }{2\pi }
\label{T1}
\end{equation}%
where surface gravity is defined as%
\begin{equation*}
\left\vert \kappa \right\vert \equiv \left. \left\langle \frac{1}{\sqrt{-h}}%
\partial _{a}\left( \sqrt{-h}h^{ab}\partial _{b}\tilde{r}\right)
\right\rangle \right\vert _{r=\left\vert r_{A}\right\vert }
\end{equation*}%
For later convention, we rewrite it
\begin{equation}
\left\vert \kappa \right\vert =\left\vert r_{A}\right\vert \left\langle
\frac{-a}{2}\left[ \frac{a}{2\dot{a}}\partial _{t}\left( \frac{\left\langle
\dot{a}\right\rangle ^{2}}{a^{2}}\right) +2\frac{\left\langle \dot{a}%
\right\rangle ^{2}}{a^{2}}\right] \right\rangle .  \label{kapa}
\end{equation}%
The projection of the 4-dimensional energy-momentum tensor is defined as%
\begin{equation*}
\left\vert T_{a}^{b}\right\vert \equiv \left\langle T_{a}^{b}\right\rangle ,
\end{equation*}%
and work density reads%
\begin{equation*}
\left\vert W\right\vert \equiv -\frac{1}{2}\left\vert T_{a}^{a}\right\vert =-%
\frac{1}{2}\left\langle T_{a}^{a}\right\rangle ,
\end{equation*}%
then the energy-supply vector is%
\begin{align*}
\left\vert \Psi _{a}\right\vert & \equiv \left. \left\langle \left\vert
T_{a}^{b}\right\vert \partial _{b}\tilde{r}+\left\vert W\right\vert \partial
_{a}\tilde{r}\right\rangle \right\vert _{r=\left\vert r_{A}\right\vert } \\
& =\left. \left\langle \left\langle T_{a}^{b}\right\rangle \partial _{b}%
\tilde{r}-\frac{1}{2}\left\langle T_{a}^{a}\right\rangle \partial _{a}\tilde{%
r}\right\rangle \right\vert _{r=\left\vert r_{A}\right\vert }.
\end{align*}%
Expressing%
\begin{equation*}
\left\vert A\right\vert \equiv 4\pi \left\vert \tilde{r}_{A}\right\vert
^{2},\;\left\vert V\right\vert \equiv \frac{4}{3}\pi \left\vert \tilde{r}%
_{A}\right\vert ^{3}
\end{equation*}%
as the area and volume of an 3-dimensional space with radius $\left\vert
\tilde{r}_{A}\right\vert $ respectively, the total energy flux on the
apparent horizon can be written as%
\begin{equation}
\left\vert \nabla E\right\vert \equiv \left\vert A\right\vert \left\vert
\Psi \right\vert +\left\vert W\right\vert \nabla \left\vert V\right\vert .
\label{dE}
\end{equation}%
For our aim, we write $\left\vert A\right\vert \left\vert \Psi \right\vert $
explicitly%
\begin{equation}
\left\vert A\right\vert \left\vert \Psi \right\vert =4\pi \left\vert \tilde{r%
}_{A}\right\vert ^{2}\left[ -\left\vert \tilde{r}_{A}\right\vert
H_{\left\langle a\right\rangle }\left( \left\langle \rho \right\rangle
+\left\langle p\right\rangle \right) dt+\frac{1}{2}\left( \left\langle \rho
\right\rangle +\left\langle p\right\rangle \right) d\left\vert \tilde{r}%
_{A}\right\vert \right] .  \label{Apsi}
\end{equation}%
Now we will consider the thermodynamics of two effective 4D reductions. }

\subsection{Thermodynamics in the low energy}

Inserting the reduced scale factor $a$ (\ref{a low energy}), we read the
surface gravity (\ref{kapa}) on the horizon (\ref{horizon}) as%
\begin{align*}
\left\vert \kappa \right\vert & =-\left\langle \frac{1}{B\cos (ky)}%
\right\rangle \frac{\left\vert \tilde{r}_{A}\right\vert }{2}\left[ \frac{1}{%
2H_{\left\langle a\right\rangle }}\partial _{t}\left( H_{\left\langle
a\right\rangle }^{2}\right) +2\left( H_{\left\langle a\right\rangle
}^{2}\right) \right] \\
& =-\frac{1}{B}\left\langle \frac{1}{\cos (ky)}\right\rangle \frac{1}{%
\left\vert \tilde{r}_{A}\right\vert }\left[ 1-\frac{\partial _{t}\left\vert
\tilde{r}_{A}\right\vert }{2H_{\left\langle a\right\rangle }\left\vert
\tilde{r}_{A}\right\vert }\right] ,
\end{align*}%
Using the energy conservation equation (\ref{ro low energy}), we can further
draw the surface gravity as%
\begin{equation}
\left\vert \kappa \right\vert =-\frac{1}{B}\left\langle \frac{1}{\cos (ky)}%
\right\rangle \frac{1}{\left\vert \tilde{r}_{A}\right\vert }\left[ 1+\frac{%
3\left( \left\langle \rho _{m}\right\rangle +\left\langle p_{m}\right\rangle
\right) \partial _{t}\left\vert \tilde{r}_{A}\right\vert }{2\left\langle
\dot{\rho}_{m}\right\rangle \left\vert \tilde{r}_{A}\right\vert }\right] ,
\label{k1}
\end{equation}%
and change $\left\vert A\right\vert \left\vert \Psi \right\vert $ (\ref{Apsi}%
) into%
\begin{align}
\left\vert A\right\vert \left\vert \Psi \right\vert & =4\pi \left\vert
\tilde{r}_{A}\right\vert ^{2}\left[ \left\vert \tilde{r}_{A}\right\vert
\frac{\left\langle \dot{\rho}_{m}\right\rangle }{3}dt+\frac{1}{2}\left(
\left\langle \rho _{m}\right\rangle +\left\langle p_{m}\right\rangle \right)
d\left\vert \tilde{r}_{A}\right\vert \right]  \notag \\
& =\frac{4\pi \left\vert \tilde{r}_{A}\right\vert ^{3}}{3}d\left\langle \rho
_{m}\right\rangle \left[ 1+\frac{3\left( \left\langle \rho _{m}\right\rangle
+\left\langle p_{m}\right\rangle \right) \partial _{t}\left\vert \tilde{r}%
_{A}\right\vert }{2\left\langle \dot{\rho}_{m}\right\rangle \left\vert
\tilde{r}_{A}\right\vert }\right] .  \label{Apsi1}
\end{align}%
To build up the first law of thermodynamics, we define entropy%
\begin{align}
S& \equiv \int \frac{\left\vert A\right\vert \left\vert \Psi \right\vert }{T}%
=\int \frac{2\pi \left\vert A\right\vert \left\vert \Psi \right\vert }{%
\kappa }  \notag \\
& =-\int \frac{1}{\left\langle \frac{1}{B\cos (ky)}\right\rangle }\frac{8\pi
^{2}\left\vert \tilde{r}_{A}\right\vert ^{4}}{3}d\left\langle \rho
_{m}\right\rangle ,  \label{S00}
\end{align}%
where the last equality has been obtained using Eqs. (\ref{k1}) and (\ref%
{Apsi1}). Substituting $\left\langle \rho _{m}\right\rangle $ (\ref{ro final}%
) and integrating Eq. (\ref{S00}), the entropy can be obtained%
\begin{equation}
S=\frac{8\pi ^{2}B}{\eta \left\langle \frac{1}{\cos (ky)}\right\rangle }%
\left[ \left( \frac{6}{\lambda }\left\vert \tilde{r}_{A}\right\vert
^{2}+r_{b}\left\vert \tilde{r}_{A}\right\vert ^{2}+r_{c}\left\vert \tilde{r}%
_{A}\right\vert ^{2}\right) +\frac{32}{3}\alpha \lambda \left\vert \tilde{r}%
_{A}\right\vert ^{2}\right] .  \label{S1}
\end{equation}%
The physical meaning of this expression is clear: The three terms in
parenthesis are contributed by the pure thin RS brane effect, the thick
brane effect, and the induced gravity effect, respectively. The remained
term describes the first order correction of GB effect. All terms are
affected by brane thickness through $B/\left\langle \eta /\cos
(ky)\right\rangle $.

At last, it should be pointed out that, using the obtained entropy
expression (\ref{S1}) and Eqs. (\ref{ro low energy}), (\ref{T1}), (\ref{dE}%
), (\ref{k1}), and (\ref{Apsi1}), one can construct the first law of
thermodynamics%
\begin{equation*}
\left\vert dE\right\vert =\left\vert T\right\vert dS+\left\vert W\right\vert
d\left\vert V\right\vert .
\end{equation*}

\subsection{Thermodynamics for a small brane thickness}

Next we are going to construct the first law in the limit for a small brane
thickness by a similar procedure of above subsection. Inserting the reduced
scale factor $a$ (\ref{a small thickness}), we obtain the surface gravity (%
\ref{kapa})%
\begin{align}
\left\vert \kappa \right\vert & =-\frac{\left\vert \tilde{r}_{A}\right\vert
}{2}\left[ \frac{\left\langle \dot{a}\right\rangle ^{2}}{\left\langle
a\right\rangle ^{2}}+\frac{\left\langle \ddot{a}\right\rangle }{\left\langle
a\right\rangle }\right] +\left\langle 1-12y^{2}\right\rangle O(\varepsilon
)+O(\varepsilon ^{2})  \notag \\
& =-\frac{\left\vert \tilde{r}_{A}\right\vert }{2}\left[ \frac{1}{%
2H_{\left\langle a\right\rangle }}\partial _{t}\left( H_{\left\langle
a\right\rangle }^{2}\right) +2\left( H_{\left\langle a\right\rangle
}^{2}\right) \right] +O(\varepsilon ^{2}).  \label{kk}
\end{align}%
Noticing that $a$ is first order in $\varepsilon $, the surface gravity
should also be read at first order in $\varepsilon $%
\begin{align*}
\left\vert \kappa \right\vert & =-\frac{\left\vert \tilde{r}_{A}\right\vert
}{2}\left[ \frac{1}{2H_{\left\langle a\right\rangle }}\partial _{t}\left(
H_{\left\langle a\right\rangle }^{2}\right) +2\left( H_{\left\langle
a\right\rangle }^{2}\right) \right] \\
& =-\frac{1}{\left\vert \tilde{r}_{A}\right\vert }\left[ 1-\frac{\partial
_{t}\left\vert \tilde{r}_{A}\right\vert }{2H_{\left\langle a\right\rangle
}\left\vert \tilde{r}_{A}\right\vert }\right] ,
\end{align*}%
where the definition of horizon (\ref{horizon}) has been used. Using the
energy conservation equation (\ref{ro small thickness}), we can further read
the surface gravity as%
\begin{equation}
\left\vert \kappa \right\vert =-\frac{1}{\left\vert \tilde{r}_{A}\right\vert
}\left[ 1+\frac{3\left( \left\langle \rho _{m}\right\rangle +\left\langle
p_{m}\right\rangle \right) \partial _{t}\left\vert \tilde{r}_{A}\right\vert
}{2\left\langle \dot{\rho}_{m}\right\rangle \left\vert \tilde{r}%
_{A}\right\vert }\right] ,  \label{k3}
\end{equation}%
and write $\left\vert A\right\vert \left\vert \Psi \right\vert $ (\ref{Apsi}%
) as%
\begin{align}
\left\vert A\right\vert \left\vert \Psi \right\vert & =4\pi \left\vert
\tilde{r}_{A}\right\vert ^{2}\left[ \left\vert \tilde{r}_{A}\right\vert
\frac{\left\langle \dot{\rho}_{m}\right\rangle }{3}dt+\frac{1}{2}\left(
\left\langle \rho _{m}\right\rangle +\left\langle p_{m}\right\rangle \right)
d\left\vert \tilde{r}_{A}\right\vert \right]  \notag \\
& =\frac{4\pi \left\vert \tilde{r}_{A}\right\vert ^{3}}{3}d\left\langle \rho
_{m}\right\rangle \left[ 1+\frac{3\left( \left\langle \rho _{m}\right\rangle
+\left\langle p_{m}\right\rangle \right) \partial _{t}\left\vert \tilde{r}%
_{A}\right\vert }{2\left\langle \dot{\rho}_{m}\right\rangle \left\vert
\tilde{r}_{A}\right\vert }\right] .  \label{Apsi2}
\end{align}%
To build up the first law of thermodynamics, we define entropy%
\begin{align}
S& \equiv \int \frac{\left\vert A\right\vert \left\vert \Psi \right\vert }{T}%
=\int \frac{2\pi \left\vert A\right\vert \left\vert \Psi \right\vert }{%
\kappa }  \label{ss} \\
& =-\int \frac{8\pi ^{2}\left\vert \tilde{r}_{A}\right\vert ^{4}}{3}%
d\left\langle \rho _{m}\right\rangle ,  \label{S20}
\end{align}%
where the last equality has been got using Eqs. (\ref{k3}) and (\ref{Apsi2}%
). Since $\left\langle \rho _{m}\right\rangle $ (\ref{roH2 small thickness})
is a function of $\tilde{r}_{A}$, we can extract the entropy in principle by
integrating Eq. (\ref{S20}). However, the exact expression is expatiatory
and we only give the entropy at first order $\alpha $, $r_{b}$, $r_{c}$
through Eq. (\ref{roH2 simplify}),
\begin{equation}
S=\frac{8\pi ^{2}}{3}\left[ \left( 2\left\vert \tilde{r}_{A}\right\vert ^{3}+%
\frac{22}{3}r_{b}\left\vert \tilde{r}_{A}\right\vert ^{2}+3r_{c}\left\vert
\tilde{r}_{A}\right\vert ^{2}\right) +48\alpha \left\vert \tilde{r}%
_{A}\right\vert -\frac{304}{9}\alpha \log (\left\vert \tilde{r}%
_{A}\right\vert )r_{b}\right] .  \label{S2}
\end{equation}%
The physical meaning of this expression is also clear: The three terms in
parenthesis are contributed by the pure thin RS brane effect, the thick
brane effect, and the induced gravity effect, respectively. The remained two
terms describe the first order GB effect, and the corresponding thickness
correction, respectively. Moreover, using the entropy expression (\ref{S20})
(or its first order approximation (\ref{S2})) and Eqs. (\ref{ro small
thickness}) (or (\ref{roH2 simplify})), (\ref{T1}), (\ref{dE}), (\ref{k3}),
and (\ref{Apsi2}), one can construct the first law of thermodynamics%
\begin{equation*}
\left\vert dE\right\vert =\left\vert T\right\vert dS+\left\vert W\right\vert
d\left\vert V\right\vert .
\end{equation*}%
\ Furthermore, comparing the entropy expressions (\ref{S1}) and (\ref{S2}),
we note following interesting features. First, all terms contribute the
entropy expression (\ref{S1}) as the form of the Bekenstein-Hawking entropy
of the 4D Einstein gravity ($\sim \left\vert \tilde{r}_{A}\right\vert ^{2}$%
). While in entropy expression (\ref{S2}), the pure thin brane effect (the
first term of (\ref{S2})) and the GB effect (the fourth term of (\ref{S2}))
embody the bulk effect \cite{Sheykhi,Sheykhi1}. This means that the RS
fine-turning relation (\ref{llamda}) and the low energy limit localize the
gravity on the brane, whatever the brane has thickness and curvature
corrections. Second, the induced gravity effect (the third terms of
parenthesis in (\ref{S1}) and in (\ref{S2})) and the small thick brane
effect (the second terms of parenthesis in (\ref{S2})) also contribute the
entropy expressions as the 4D Bekenstein-Hawking entropy. It implies that
both effects are localized on four dimensions.

\section{Conclusion and discussion}

In this paper, we have generalized the thick RS braneworld scenario
presented in \cite{Mounaix} in the presence of a 4D scalar curvature from
induced gravity on the brane, and a 5D GB curvature term. We have obtained
the generalized Friedmann equation governing the cosmological evolution
inside the thick brane. As discussed in RS brane world \cite{Mounaix}, it is
a very instructive because the generalized equation interpolates between the
familiar 5D GB gravity and the more attractive DGP brane gravity with GB
correction. This can be seen clearly from the Eq. (\ref{jun}) for DGP brane
cosmology with GB correction when the brane thickness is infinitely thin.
For infinitely thick brane, the Kaluza-Klein picture where matte is
homogeneously distributed over the extra-dimension and the ordinary
Friedmann equations are recovered, is not affected by the induced gravity
correction indeed, see Eq. (\ref{jun1}). We hope that our generalized
Friedmann equation will help in the intuitive understanding of the
unconventional DGP brane gravity with GB correction, and clarify the
delimitations of the various regimes where different Friedmann equations
have to be applied. Moreover, we have proposed the generalized version of
the thick RS fine-turning condition between the bulk cosmological constant
and the brane tension. In our case, this cancellation condition depends
explicitly on curvature correction besides the brane thickness and yields
back the familiar condition without the curvature correction and in the thin
brane limit. Using this fine-turning condition and assuming the cosmological
matter content of the brane is small with respect to its tensor, we have got
the effective 4D reduced cosmological equations. It should be noticed that
the GB effect is assumed to be small. In the limit where the brane thickness
is small, we have found another effective reduced cosmological equations,
whereas the induced gravity effect is assumed to be small. Whether the
restriction of small curvature corrections can be released is worth to be
further explored.

Furthermore, we have studied the thermodynamics of generalized thick
braneworld. Similar to the definition of 4D cosmological quantities, the 4D
thermodynamic quantities are also defined by integrating over fifth
coordinate. It has been shown that two effective 4D reductions of the
generalized Friedmann equation can be written directly in the form of the
first law of thermodynamics on the apparent horizon. We stress that this
result is not intuitive because we do not know previously the reduced
Friedmann equations and reduced effective 4D thermodynamics brane
quantities. In particular, one should notice that we have omitted the
higher-order terms in the reduced thermodynamics brane quantities (\ref{kk}%
), otherwise we can not extract the entropy from the first law of
thermodynamics, see Eq. (\ref{ss}). This result strongly shows that the
connection between gravity and thermodynamics is not an accident result. For
the generalized Friedmann equation (\ref{general FM}) and the conservation
equation (\ref{ro}), however, we did not obtain the thermodynamics
accordingly. In fact, in \cite{Mounaix}, it has been shown that if the brane
thickness is not small enough, the effective 4D cosmological equations can
be obtained but in the price of introducing auxiliary quantities. Observing
the Eqs. (\ref{S00}) (\ref{S20}), one can find that the reason why we can
extract the entropy expressions is that $\left\langle \rho _{m}\right\rangle
$ (\ref{ro final}) (\ref{roH2 small thickness}) are the functions of $\tilde{%
r}_{A}$ only. If the Friedmann equations have auxiliary quantities, $%
\left\langle \rho _{m}\right\rangle $ will depend on them and then we can
not extract the entropy in general by integrating Eqs. (\ref{S00}) (\ref{S20}%
). This is a sign that the full 5D description is more adequate, and the 4D
thermodynamics can not reflect the gravity of five dimensions. On the other
hand, we can speculate that thermodynamics can not reflect the gravity if
which has some auxiliary quantities. We emphasize that our study is
restricted on equilibrium thermodynamics, whether the generalized Friedmann
equations may be described by non-equilibrium thermodynamics is not clear in
present work. At last, we would like to point out that our speculation is
consonant with the result that no equilibrium thermodynamics is constructed
for nonlinear gravity and scalar-tensor gravity, because there are
nontrivial auxiliary quantities $df/dR$ \cite{Eling,Akbar} and $F(\Phi )$
\cite{Cao} for nonlinear $f(R)$ gravity and scalar-tensor gravity
respectively. We will further investigate whether the auxiliary quantities
are essential in the relation between the gravity and thermodynamics in
other work \cite{Wu2}.

\section*{Acknowledgment}

S. F. Wu wishes to thank Y. X. Liu for interesting discussion. This work was
supported by NSFC under Grant Nos. 10575068 and 10604024, the Shanghai
Research Foundation No. 07dz22020, the CAS Knowledge Innovation Project Nos.
KJcx.syw.N2, the Shanghai Education Development Foundation, and the
Innovation Foundation of Shanghai University.

\end{document}